# Superconductivity with a Violation of Pauli Limit and Evidences for Multigap in $\eta$-Carbide type Ti$_4$Ir$_2$O

Bin-Bin **Ruan**$^{a,b,*}$, Meng-Hu **Zhou**$^{a,b}$, Qing-Song **Yang**$^{b,c}$, Ya-Dong **Gu**$^{b,c}$, Ming-Wei **Ma**$^{b}$, Gen-Fu **Chen**$^{b,c}$ and Zhi-An **Ren**$^{b,c,*}$

$^a$*Songshan Lake Materials Laboratory, Dongguan, Guangdong 523808, China*

$^b$*Institute of Physics and Beijing National Laboratory for Condensed Matter Physics, Chinese Academy of Sciences, Beijing 100190, China*

$^c$*School of Physical Sciences, University of Chinese Academy of Sciences, Beijing 100049, China*

---

## ABSTRACT

We report the synthesis, crystal structure, and superconductivity of Ti$_4$Ir$_2$O. The title compound crystallizes in an $\eta$-carbide type structure of the space group $Fd\bar{3}m$ (No. 227), with lattice parameters $a = b = c = 11.6194(1)$ Å. The superconducting $T_c$ is found to be 5.1 $\sim$ 5.7 K. Most surprisingly, Ti$_4$Ir$_2$O hosts an upper critical field of 16.45 T, which is far beyond the Pauli paramagnetic limit. Strong coupled superconductivity with evidences for multigap is revealed by the measurements of heat capacity and upper critical field. First-principles calculations suggest that the density of states near the Fermi level originates from the hybridization of Ti-$3d$ and Ir-$5d$ orbitals, and the effect of spin-orbit coupling on the Fermi surfaces is prominent. Large values of the Wilson ratio ($R_W \sim$ 3.9), the Kadowaki-Woods ratio ($A/\gamma^2 \sim 9.0 \times 10^{-6}$ $\mu\Omega$ cm/(mJ mol$^{-1}$ K$^{-1}$)$^2$), and the Sommerfeld coefficient ($\gamma = 33.74$ mJ mol$^{-1}$ K$^{-2}$) all suggest strong electron correlations (similar to heavy fermion systems) in Ti$_4$Ir$_2$O. The violation of Pauli limit is possibly due to a combination of strong-coupled superconductivity, large spin-orbit scattering, and electron correlation. With these intriguing behaviors, Ti$_4$Ir$_2$O serves as a candidate for unconventional superconductor.

---

## 1. Introduction

For their superior abilities to carry large currents, superconductors have been widely used to generate high magnetic fields at different settings, such as medical magnetic resonance imaging (MRI), maglevs, particle accelerators, *etc.*[1, 2]. However, applications to these uses have been limited to only a few superconducting compounds, primarily NbTi or Nb$_3$Sn[3]. Apart from the superconducting transition temperature ($T_c$), the critical current density, and the ductibility, one major limitation comes from the upper critical field ($H_{c2}$) in a type-II superconductor.

At low temperatures, the Zeeman energy splitting ($\Delta E$) is the key factor limiting $H_{c2}$. When the external magnetic field is large enough ($\geq \mu_0 H_P$), driving $\Delta E$ comparable with twice of the superconducting energy gap, the Cooper pairs are broken. In a Bardeen–Cooper–Schrieffer (BCS) weak coupling scenario, the Pauli paramagnetic limit (or, Pauli limit for short) is expressed by $\mu_0 H_P = 1.86 \times T_c$ (T/K).

$H_{c2}$ of most type-II superconductors falls below $H_P$, regardless of the value of $T_c$[4]. While a violation of the Pauli limit is often regarded as a side evidence for unconventional superconductivity, as in heavy fermions[5], iron-based superconductors[6], non-centrosymmetric superconductors[7, 8], and more recently, in $A_2$Cr$_3$As$_3$ ($A$ = Na, K, Rb, Cs)[9, 10, 11] or magic-angle twisted trilayer graphene[12].

In addition to unconventional superconductivity, large spin-orbit scattering[13], significant electron correlations, strong electron-phonon coupling[14], or a highly anisotropic structure[15] all tend to increase $H_{c2}$, pushing it beyond $H_P$. Nevertheless, superconductors with a violation of Pauli limit are still very rare.

In this study, we focus on a suboxide of the so-called "$\eta$-carbide" family, also known as the E9$_3$-type phases. The $\eta$-carbide type phase adopts a cubic crystal structure of the space group $Fd\bar{3}m$ (No. 227), which is derived from the Ti$_2$Ni type after small atoms, typically C, N, or O, take the interstitial sites. W$_3$Fe$_3$C, the first member in the family, was discovered in the 1950's[16, 17]. Later on, a large number of carbides, nitrides, and oxides of the same structural family have been reported. Depending on different crystallographic sites the atoms take, there are four stoichiometries of $\eta$-carbide type phases, namely (T and T' stand for transition metals, X = C, N, O): (a) T$_4$T'$_2$X, as in Mo$_4$Co$_2$C, Zr$_4$Re$_2$O[17];(b) T$_3$T'$_3$X, as in W$_3$Fe$_3$C, Nb$_3$Cr$_3$N[16];(c) T$_6$T'$_6$X, as in Mo$_6$Ni$_6$C[18]; (d) T$_8$T'$_4$X, as in Nb$_8$Zn$_4$C[19]. Although they have been discovered for more than half a century, reports on these phases mainly focus on their hardness[20], the hydrogen-sorption[21], or the catalytic abilities[22], few of them has been examined for their electronic properties.

Ku *et al.* reported several $\eta$-carbide type superconductors in 1984[23], including Ti$_4$Co$_2$O ($T_c$ = 3.1 K), Zr$_4$Os$_2$O ($T_c$ = 3.0 K), Nb$_4$Rh$_2$C ($T_c$ = 8.9 K), *etc.*. Recently, superconductivity in Zr$_4$Rh$_2$O$_{0.6}$[24] and Nb$_4$Rh$_2$C$_{1-\delta}$[25] was systematically studied. Nb$_4$Rh$_2$C$_{1-\delta}$ was found to host a large $\mu_0 H_{c2}(0)$ of 28.5 T, violating the Pauli paramagnetic limit. This is not trivial since $\eta$-carbide type phases are not highly anisotropic in their crystal structures. Thus, strong electron correlations, or even unconventional superconductivity can be expected in the superconductors of $\eta$-carbide

---

*Corresponding authors

✉ bbruan@mail.ustc.edu.cn (B. Ruan); renzhian@iphy.ac.cn (Z. Ren)

ORCID(s): 0000-0003-4642-7782 (B. Ruan); 0000-0003-4308-7372 (Z. Ren)

type. However, evidences for these speculations are still absent.

Here we report the synthesis and a systematical study of superconductivity in $Ti_4Ir_2O$. Bulk superconductivity of $T_c$ = 5.12 K is revealed by magnetic susceptibility and heat capacity measurements, confirming a previous report by Matthias *et al.* in 1963[26]. In that paper, superconductivity in "$Ti_{0.573}Ir_{0.287}O_{0.14}$" was reported without further details. Notice that such an early study needs reexamination. For example, superconductivity of $T_c$ = 11.8 K in "$Zr_{0.61}Rh_{0.285}O_{0.105}$" reported in the same paper[26] was later ascribed to $CuAl_2$-type $Zr_2Rh$[27]. In addition, superconducting parameters of $Ti_4Ir_2O$ are determined in our study, among which the $\mu_0 H_{c2}(0)$ is 16.45 T, far above the Pauli limit (9.52 T). Moreover, the deviation of $\mu_0 H_{c2}(T)$ from the Werthamer–Helfand–Hohenberg (WHH) theory, and the heat capacity in the superconducting state suggest a multigap nature. Large electron correlations in $Ti_4Ir_2O$ are evidenced by both experimental results and first-principles calculations.

## 2. Methods

Polycrystalline samples of $Ti_4Ir_2O$ were prepared by solid state reactions. Stoichiometric amount of titanium (powder, 99.99%), iridium (powder, 99.99%), and $TiO_2$ (powder, 99.95%) was mixed thoroughly, and then cold pressed into pellets. The pellets were placed in corundum crucibles, which were subsequently sealed into niobium tubes under purified argon. The niobium tubes were heated to 1800 K, held at the temperature for 20 hours, and then cooled down to room temperature by switching off the furnace. To avoid the oxidation of niobium tubes at high temperatures, heating treatments were carried out under an argon atmosphere. Grey and dense solid was obtained after opening the niobium tubes. No reaction between the samples and the niobium tubes was observed, and the weight losses after the reactions were negligible. We also performed reactions at lower temperatures or using arc-melting methods. However, these alternative methods did not produce single-phase samples. (see Supporting Information)

Powder x-ray diffraction (XRD) measurements were carried out on a PAN-analytical x-ray diffractometer equipped with Cu-K$\alpha$ radiation at room temperature. The Rietveld refinement of the XRD results were performed with the GSAS package[28]. Temperature dependent electrical resistivity and heat capacity of the sample was measured on a Quantum Design physical property measurement system (PPMS). DC magnetization was measured on a Quantum Design magnetic property measurement system (MPMS). Details for the physical property measurements can be found elsewhere[29]. All the data from the magnetization measurements were corrected based on the demagnetization factor, which was determined by the dimensions of the rectangular sample[30].

First-principles calculations were carried out in the frame of the density functional theory (DFT), using the QUANTUM ESPRESSO (QE) package[31, 32, 33]. Generalized gradient approximation (GGA) exchange-correlation functionals of PBEsol[34] were selected, together with the projector augmented wave pseudopotentials from the PSLIBRARY package[35]. To take spin-orbit coupling (SOC) into account, we performed both the scalar-relativistic and fully-relativistic calculations. Energy cut-offs were 100 Ry and 1000 Ry for the wavefunctions and the charge densities, respectively. A Monkhorst-Pack grid of $8^3$ $k$-points was used for the self-consistent calculation, and a denser grid of $12^3$ $k$-points was used to generate the density of states (DOS). Before the self-consistent calculation, the cell parameters, as well as the atomic positions, were fully relaxed starting from the experimental values, using the Broyden–Fletcher–Goldfarb–Shanno (BFGS) algorithm. The grid of $k$-mesh was checked for convergence. The Barder analysis was performed with the help of the CRITIC2 package[36]. Fermi surfaces were constructed on a superfine grid of $101^3$ $k$-points, with the help of maximally localized Wannier functions (MLWFs) generated by the WANNIER90 code[37, 38].

## 3. Results

Figure 1 displays the powder XRD pattern of $Ti_4Ir_2O$ collected at room temperature. We find that $Ti_4Ir_2O$ crystallizes in an $\eta$-carbide type structure, which is of the space group $Fd\bar{3}m$ (No. 227). In the structure, Ti and Ir form a $Ti_2Ni$-type framework, where O takes the 16$c$ (distorted octahedral) interstitial sites (see the inset of Figure 1). We notice that the inclusion of oxygen is key to stablize the $Ti_4Ir_2O$ phase, since no $Ti_2Ni$-type "$Ti_2Ir$" phase can be acquired

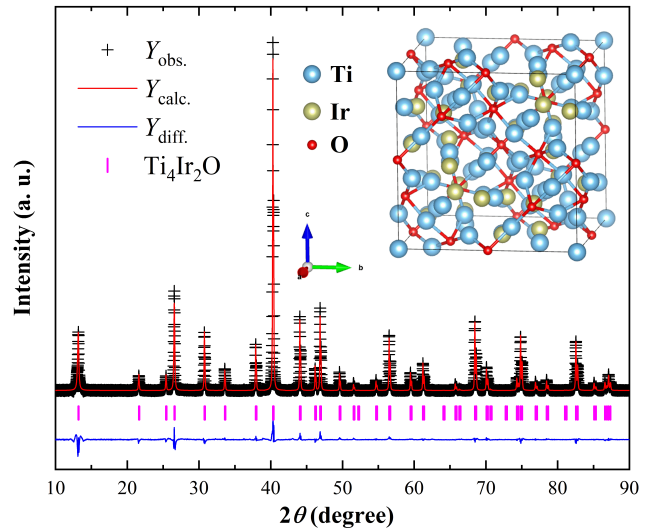

**Figure 1:** Room-temperature XRD pattern of $Ti_4Ir_2O$ and its Rietveld refinements. $Y_{obs.}$, $Y_{calc.}$, and $Y_{diff.}$ stand for the observed intensity, the calculated intensity, and the difference between, respectively. The vertical bars indicate the Bragg positions for $\eta$-carbide type $Ti_4Ir_2O$. The conventional unit cell is shown as the inset.

**Table 1**

Crystallographic parameters of Ti₄Ir₂O from Rietveld Refinement[a] of room temperature XRD. Parameters from first-principles calculations were also listed for comparison.

| Space Group | | $Fd\bar{3}m$ (No. 227) | | |
|---|---|---|---|---|
| $a$ (Å) | | 11.6194(1) | | |
| $a_{w/o-SOC}$ (Å) | | 11.5562 | | |
| $a_{with-SOC}$ (Å) | | 11.5531 | | |
| Atom (position) | $x$[b] | $U_{eq}$(0.01Å²)[c] | $x_{w/o-SOC}$ | $x_{with-SOC}$ |
| Ti1 (16$c$) | 0 | 0.348 | 0 | 0 |
| Ti2 (48$f$) | 0.4412 | 0.065 | 0.4428 | 0.4419 |
| Ir (32$e$) | 0.2145 | 0.276 | 0.2160 | 0.2159 |
| O (16$d$) | 0.5 | 0.633 | 0.5 | 0.5 |

[a] $R_p = 5.80\%$, $R_{wp} = 3.49\%$.

[b] $y_{Ti1} = z_{Ti1} = 0$, $y_{Ti2} = z_{Ti2} = 0.125$, $x_{Ir} = y_{Ir} = z_{Ir}$, $x_O = y_O = z_O$.

[c] $U_{eq}$ is defined as one-third of the trace of the orthogonalized $U_{ij}$ tensor.

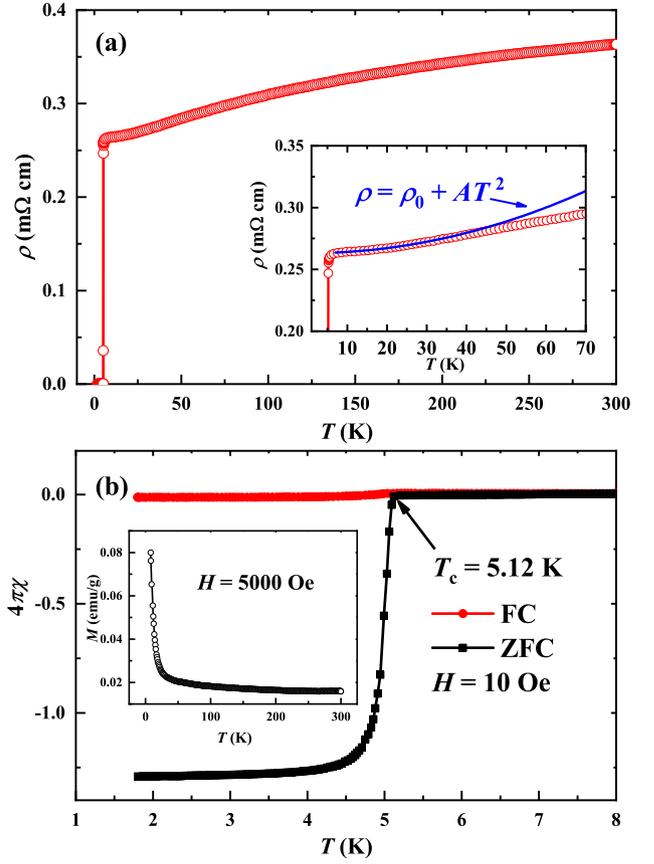

**Figure 2:** (a) Temperature dependence of resistivity of Ti₄Ir₂O under zero magnetic field. Normal state data below 35 K can be described by equation(1), as shown in the inset. (b) Temperature dependence of DC magnetic susceptibility of Ti₄Ir₂O at low temperatures. Inset shows the temperature dependence of the normal state magnetization from 8 K to 300 K.

(nor has it been reported in the Ti–Ir phase diagram[39]). The crucial roles of interstitial C, N, or O have been frequently observed in other η-carbide type phases, such as W₃Fe₃C[17], Zr₄Pd₂N[23], or Zr₄Rh₂O[24].

Rietveld refinement was performed and the result is shown in Figure 1, giving cell parameters $a = b = c = 11.6194(1)$ Å. The value is close to the reference one 11.620 Å[26]. The occupancy on each site is refined to be 1.0 (except for that of oxygen, which is fixed to 1. We will come back to this issue in Section 4). Detailed refined crystallographic parameters are summarized in Table 1. The fully relaxed cell parameters and atomic positions are also listed for comparison. Notice that the discrepancies between the DFT values and the experimental values are within 0.6%, and the inclusion of SOC has negligible effects on the crystallographic parameters.

Superconductivity is evidenced in both resistivity and magnetization measurements, as shown in Figure 2. At the normal state, Ti₄Ir₂O displays a metallic behavior, with the resistivity ($\rho$) decreases upon cooling. The residual resistivity ratio ($RRR = \rho(300 \text{ K})/\rho(6 \text{ K}) = 1.39$) is quite small, which is typical for a transition metal suboxide (for example, see [24, 40]). The normal state $\rho(T)$ shows a convex feature ("saturation") above 50 K. This can be explained with a parallel-resistor model, in which a strong electron-phonon coupling takes place (see, for instance, the A15 superconductors[41, 42] and SrPt₃P[43]). Alternatively, it can be interpreted by an inter-band scattering, known as the Bloch–Grüneisen–Mott (BGM) mechanism[29, 44, 45]. The inset of Figure 2 shows $\rho(T)$ below 70 K, in which we find that the data below 35 K following a Fermi liquid behavior:

$$\rho(T) = \rho_0 + AT^2, \tag{1}$$

giving $\rho_0 = 0.263$ mΩ cm, $A = 10.25$ nΩ cm K⁻². At lower temperatures, $\rho(T)$ starts to drop from 5.22 K ($T_c^{onset}$), reaching zero at 5.13 K ($T_c^{zero}$), indicating the occurrence of superconductivity. Such a sharp transition implies a superior homogeneity of our sample.

In Figure 2(b), we show the DC magnetic susceptibility ($\chi$) of Ti₄Ir₂O from 1.8 K to 8.0 K. The large diamagnetic signal in the zero-field cooling (ZFC) process confirms the bulk nature of the superconductivity. The field cooling (FC) gives a much smaller diamagnetism, suggesting a substantial pinning effect in the sample. $T_c$ determined from the magnetization measurement is 5.12 K, which is in very good agreement with the value from $\rho(T)$.

Figure 3(a) shows the superconducting transition under various magnetic fields. Clearly, the transition is suppressed to lower temperatures when applying external fields. However, the suppression of $T_c$ is very slow. For instance, under a magnetic field of 9 T, $T_c$ is still above 2.5 K. Temperature dependence of the heat capacity ($C_p$) shows a distinct anomaly at low temperatures (Figure 3(d)), which again validates the bulk superconductivity. $T_c$ can be determined from the $C_p/T$–$T$ curves based on an entropy balance method. At zero magnetic field, $T_c$ determined from $C_p$ is 5.12 K, which is in excellent agreement with the values from $\rho(T)$ and $\chi(T)$.

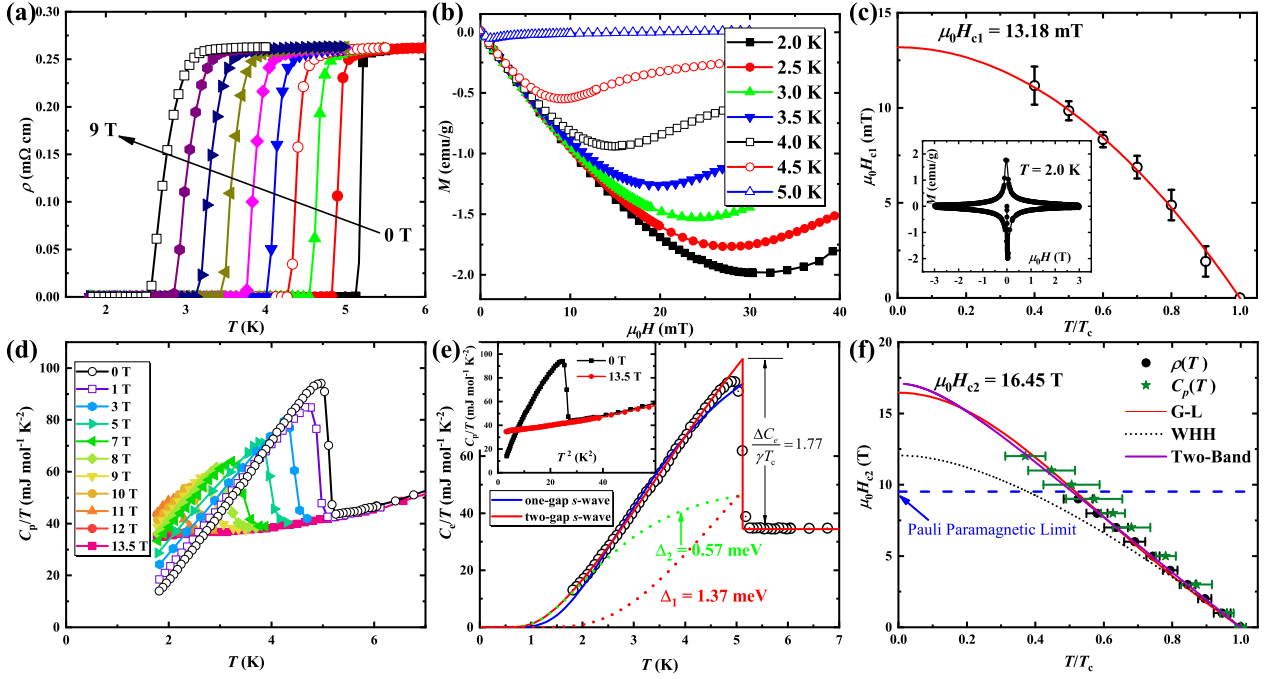

**Figure 3:** (a) Superconducting transition on $\rho(T)$ under magnetic fields from 0 T to 9 T, the field increment between each curve is 1 T. (b) Isothermal magnetization under different temperatures. (c) Lower critical fields ($H_{c1}$) under different temperatures, and the fit according to the G–L relation. Inset shows the isothermal magnetization loop at 2.0 K. (d) Low temperature heat capacity under magnetic fields from 0 T to 13.5 T. (e) Electronic contribution of heat capacity at zero magnetic field. Solid curves show fits based on a one-gap $s$-wave model (blue), and a two-gap $s$-wave model (red). Contributions from the two gaps are displayed as dotted curves. (f) Upper critical fields ($H_{c2}$) under different temperatures, determined from $\rho(T)$ and $C_p(T)$. The fitting curves are based on the $\rho(T)$ data, according to the G–L (red), WHH(dotted), and two-band (purple) model, respectively. The value of $\mu_0 H_{c2}(0)$ is from the G–L fit.

Results from $\rho(T)$ and $C_p/T$–$T$ under various magnetic fields allow us to determine $H_{c2}$ of Ti₄Ir₂O. In Figure 3(f), the points from the $\rho(T)$ measurements are determined from the midpoints of the superconducting transitions, while points from $C_p/T$–$T$ curves are determined using the method described above, error bars correspond to the transition width. The data from $\rho(T)$ and $C_p/T$–$T$ agree well with each other, while the transition widths are larger in the $C_p$ measurements. The Pauli paramagnetic limit for a superconductor with $T_c = 5.12$ K is $\sim 9.52$ T. However, as shown in Figure 3(f), superconductivity still survives when the magnetic field is 12 T. This conclusively indicates a violation of Pauli limit in Ti₄Ir₂O.

The fit to $H_{c2}$ is carried out using the WHH relation[46]:

$$\ln(\frac{1}{t}) = (\frac{1}{2} + \frac{i\lambda_{SO}}{4\gamma_w})\psi(\frac{1}{2} + \frac{h + i\gamma_w + \lambda_{SO}/2}{2t})$$
$$+ (\frac{1}{2} - \frac{i\lambda_{SO}}{4\gamma_w})\psi(\frac{1}{2} + \frac{h - i\gamma_w + \lambda_{SO}/2}{2t}) - \psi(\frac{1}{2}),$$
$$(2)$$

in which $t = T/T_c$ is the normalized temperature, $h = -(4/\pi^2)H_{c2}(T)/[T_c(dH_{c2}/dT)|_{T=T_c}]$ is the dimensionless magnetic field, and $\psi(x)$ is the digamma function. $\gamma_w =$

$(\alpha^2 h^2 - \lambda_{SO}^2/4)^{1/2}$, where $\alpha$ is the Maki parameter, and $\lambda_{SO}$ is the parameter measuring the spin-orbit scattering. WHH fit of $H_{c2}$ gives: $\alpha = 1.90$, $\lambda_{SO} = 14.05$. Such a large $\lambda_{SO}$ means that there is little Pauli limiting effect in Ti₄Ir₂O, and the spin-orbit scattering dominates. However, as shown in Figure 3(f), it fails to reproduce the $H_{c2}$ data under magnetic fields beyond 7 T, yielding an underestimated $\mu_0 H_{c2}(0)$. In fact, $\lambda_{SO} \gg 1$ also implies that WHH theory, which assumes a small $\lambda_{SO}$, is not suitable in our case. The failure of WHH fit comes from the linearity feature of the temperature dependence of $H_{c2}$. Such a deviation can be explained by multigap superconductivity, as observed in MgB₂[47], 2$H$-NbSe₂[48], iron-based superconductors[6], and YRe₂SiC[49].Indeed, the $H_{c2}$ data is excellently reproduced by a two-band model developed by Gurevich[50], as indicated by the purple line in Figure 3(f). This is not surprising because the two-band model is based on many fitting parameters. To gain insights into the nature of the intra-band and inter-band scattering, detailed studies such as nuclear magnetic resonance (NMR) or muon spectroscopy are needed.

Alternatively, $H_{c2}$ can also be fitted with the phenomenological Ginzburg–Landau (G–L) relation: $\mu_0 H_{c2}(T) = \mu_0 H_{c2}(0)\frac{1-(T/T_c)^2}{1+(T/T_c)^2}$. The G–L fit is also shown in Figure 3(f), giving $\mu_0 H_{c2}(0) = 16.45$ T, which is much larger than the Pauli paramagnetic limit.

The isothermal magnetization $(M-H)$ curves, shown in Figure 3(b), give us knowledge about the lower critical field $(H_{c1})$. The magnetic field where the $M-H$ curve starts to depart from the initial Meissner states is determined as $H_{c1}$, which is plotted in Figure 3(c). A G–L fit is carried out with the formula: $\mu_0 H_{c1}(T) = \mu_0 H_{c1}(0)[1-(T/T_c)^2]$, giving $\mu_0 H_{c1}(0) = 13.18$ mT. A full magnetization loop at 2.0 K is shown as the inset of Figure 3(c), indicating type-II superconductivity in Ti₄Ir₂O.

Based on the value of $H_{c1}(0)$ and $H_{c2}(0)$, a bunch of superconducting parameters, such as the G–L coherence length $(\xi_{GL})$, the penetration depth $(\lambda_{GL})$, the G–L parameter $(\kappa_{GL})$, and the thermodynamic field $(H_c(0))$ can be estimated. These parameters are summarized in Table 2. Details for calculating the parameters can be found in our previous study[29]. Notice that the value of $\kappa_{GL}$ is close to the isostructural superconductor Nb₄Rh₂C₁₋δ[25], and is far beyond $1/\sqrt{2}$, suggesting an extreme type-II character.

Inset of Figure 3(e) shows the heat capacity under zero field, and under a field of 13.5 T, which totally suppress the superconducting transition above 1.8 K. The normal state data can be fitted with Debye model: $C_p(T) = \gamma T + \beta T^3 + \delta T^5$, in which the three terms are the electronic contribution to the heat capacity $(C_e)$, the harmonic phonon contribution, and the anharmonic phonon contribution, respectively. The fitting parameters are: $\gamma = 33.74$ mJ mol⁻¹ K⁻², $\beta = 0.265$ mJ mol⁻¹ K⁻⁴, and $\delta = 2.53$ μJ mol⁻¹ K⁻⁶. Debye temperature $(\Theta_D)$ can be obtained by:

$$\Theta_D = \left(\frac{12\pi^4 N R}{5\beta}\right)^{1/3}, \tag{3}$$

in which $N$ is the number of atoms per formula, and $R$ is the ideal gas constant. From equation(3) we get $\Theta_D = 372$ K.

According to the McMillan relation[51], the electron-phonon coupling strength $(\lambda_{ep})$ can be expressed by:

$$\lambda_{ep} = \frac{1.04 + \mu^* \ln(\Theta_D/1.45 T_c)}{(1-0.62\mu^*)\ln(\Theta_D/1.45 T_c) - 1.04}. \tag{4}$$

Setting the Coulomb screening parameter $\mu^*$ to 0.13, a typical value for intermetallics, we get $\lambda_{ep} = 0.61$, suggesting moderate coupling in Ti₄Ir₂O. The density of states (DOS) at the Fermi level $(E_F)$ is calculated with $N(E_F) = 3\gamma/[\pi^2 k_B^2(1+\lambda_{ep})] = 8.83$ eV⁻¹ per formula unit (f. u.).

$C_e$ gives us more insights about the nature of the superconductivity. To exclude the contribution of any possible impurities, $C_e$ is obtained by subtracting $C_p|_{\mu_0 H=0} + \gamma$ on $C_p|_{\mu_0 H=13.5\,\text{T}}$. The result is shown in Figure 3(e). At the superconducting state, $C_e$ quickly approaches zero, indicating a nodeless feature of the superconducting gap. Therefore, fit to the $C_e$ is carried out based on isotropic $s$-wave models. However, a one-gap $s$-wave model fails to describe $C_e$ at the superconducting state. Entropy balance at the superconducting transition is not fulfilled either. On the other hand, a two-gap $s$-wave model perfectly describes the $C_e$ behavior, while maintaining the entropy balance. The two superconducting gaps are determined to be $\Delta_1 = 1.37$ meV, and $\Delta_2 = 0.57$ meV. Notice that the larger one is above the BCS weak coupling value $(\Delta_1/k_B T_c = 3.10 > 1.76)$. This again indicates an enhanced electron-phonon coupling in Ti₄Ir₂O, which is further evidenced by the large $C_e$ jump at $T_c$ $(\Delta C_e/\gamma T_c = 1.77)$.

**Table 2**
Normal state and superconducting parameters of Ti₄Ir₂O.

| Parameter (unit) | Value | Notes |
|---|---|---|
| $T_c^{onset}$ (K) | 5.22 | from $\rho(T)$. $T_c$ ranges from 5.1 K to 5.7 K in samples prepared with different methods, see Supporting Information. |
| $T_c^{zero}$ (K) | 5.13 | |
| $T_c^{mag.}$ (K) | 5.12 | from $\chi(T)$. |
| $T_c^{CHC}$ (K) | 5.12 | from $C_e(T)$. |
| $\mu_0 H_{c1}(0)$ (mT) | 13.18 | from G–L fit. |
| $\mu_0 H_{c2}(0)$ (T) | 16.45 | from G–L fit. |
| $\mu_0 H_c(0)$ (T) | 0.23 | |
| $\xi_{GL}$ (nm) | 4.47 | |
| $\lambda_{GL}$ (nm) | 236.1 | |
| $\kappa_{GL}$ | 52.8 | |
| $\gamma$ (mJ mol⁻¹ K⁻²) | 33.74 | |
| $\beta$ (mJ mol⁻¹ K⁻⁴) | 0.265 | |
| $\Theta_D$ (K) | 372 | |
| $\lambda_{ep}$ | 0.61 | |
| $\Delta C_e/\gamma T_c$ | 1.77 | |
| $\Delta_1/k_B T_c$ | 3.10 | |
| $\Delta_2/k_B T_c$ | 1.29 | |
| $N(E_F)$ (eV⁻¹ per f. u.) | 8.83 | experimental value deduced from $\gamma$. |
| $N'(E_F)$ (eV⁻¹ per f. u.) | 7.26 | theoretical value from DFT calculation with SOC. |

Results from the first principles calculation are shown in Figure 4, with electronic band structures in Figure 4(a) and DOS in Figure 4(b) and (c). Both the scalar relativistic (w/o SOC) and the fully relativistic (with SOC) results are shown. In both cases, there are five bands crossing the Fermi level, consistent with the metallic nature of Ti₄Ir₂O. The existence of multiple bands crossing $E_F$ also makes multigap superconductivity plausible. Notice that the energy dispersion near $E_F$ is dramatically changed when SOC is turned on. For example, band splitting occurs along the Γ–X, Γ–K, and L–W lines. This makes the Fermi surface topology very different when considering SOC (see Figure S5). Most absorbingly, nesting between the electron-like and hole-like pockets becomes prominent when SOC is turned on.

The Barder analysis was based on the converged charge densities, resulting in a valence configuration of Ti^{1.32+}₄ Ir^{1.94−}₂O^{1.40−}. This means charge transferring from Ti to Ir and O. The charge density and electron localization function (ELF) maps are shown in Figure S4, from which ionic Ti-O bonds and metallic Ti-Ir bonds can be identified.

From the projected DOS, shown in Figure 4(b) and (c), we notice the states from O are absent near $E_F$. While they hybridize with states from Ti at ~ −5.6 to −2 eV, indicating the formation of Ti-O bonds. DOS at $E_F$ is slightly lowered with SOC, giving a theoretical $N'(E_F) = 7.26$ eV$^{-1}$ per f. u.. This value is comparable with the one calculated from $\gamma$. The slightly enhancement of experimental $N(E_F)$ is possibly due to the electron-electron correlations. Notice that $E_F$ locates near a pronounced peak of DOS, which arises from the relatively flat energy dispersion of the bands near $E_F$. After the projection to each atomic orbital, we find that DOS at $E_F$ originates mainly from the Ti-3$d$ orbitals, with minor contributions from the Ir-5$d$ components.

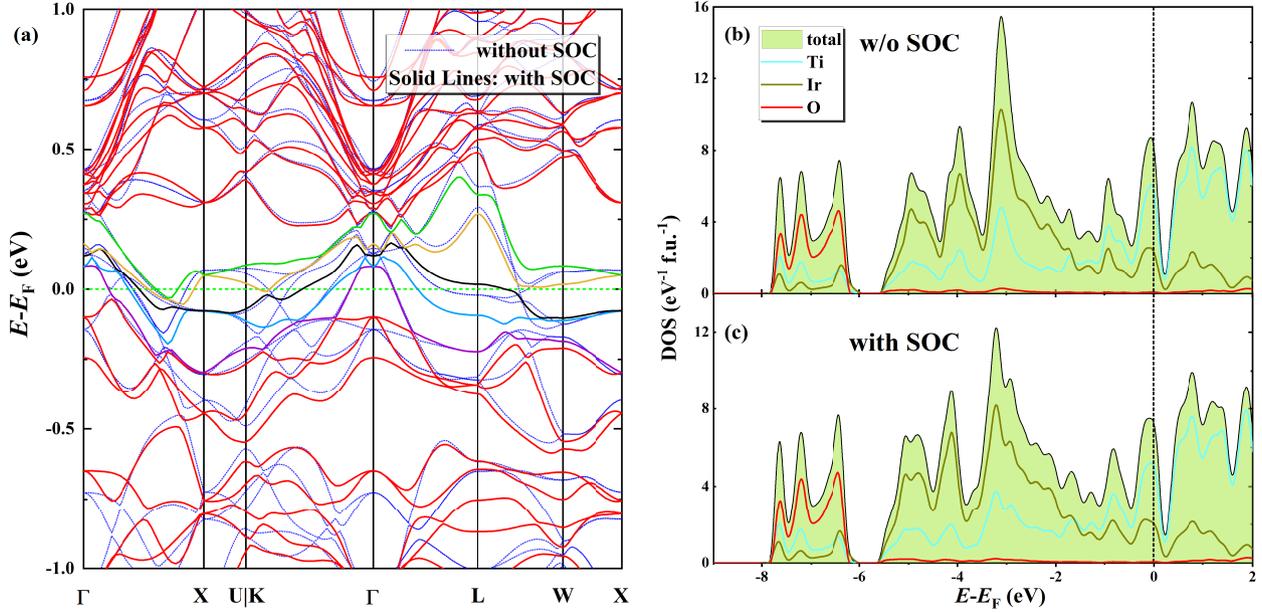

**Figure 4:** (a) Electronic band structures of Ti$_4$Ir$_2$O from DFT calculation. Dotted lines show the results without SOC, while the SOC results are shown as solid lines. Solid lines in colors other than red indicate the bands crossing the Fermi level. (b) Projected DOS without SOC. (c) Projected DOS with SOC.

## 4. Discussion and conclusion

First we would like to discuss the stoichiometry of Ti$_4$Ir$_2$O, because $\eta$-carbide type phases frequently host vacancies on the interstitial sites[24, 25]. According to the preparation procedure and the XRD results, however, the oxygen vacancies should be very close to zero. On the other hand, there should be a tiny amount of oxygen vacancies, since samples prepared at 1250 K were found to host smaller lattices and an enhanced $T_c$ of 5.7 K (see Supporting Information). In $\eta$-carbide type suboxides, the lattice shrinks when more oxygen is incorporated, as observed in Zr$_4$Rh$_2$O$_{0.6}$[24]. This means that the oxygen contents in the 5.7 K samples are larger. The change of $T_c$ should be attributed to the slight non-stoichiometry of Ti$_4$Ir$_2$O, given that DOS has a steep slope at $E_F$. In other words, a small amount of electron doping can dramatically change the value of $N(E_F)$. For example, $N(E_F)$ would be reduced to 6.80 eV$^{-1}$ per f. u. if 10% oxygen vacancies were introduced (assuming a rigid band model). The evolution of $T_c$ upon vacancies requires detailed doping studies in the future.

Next we focus on the origin of the large $H_{c2}$. One possible cause of the enhancement of $H_{c2}$ is the spin-triplet component in the order parameter, which can be induced by non-centrosymmetric crystal structures[7, 8]. However, this should not be the case since Ti$_4$Ir$_2$O hosts a centrosymmetric crystal structure. Another possibility of enhancement due to anisotropy, as observed in many low dimensional superconductors, such as Li$_{0.9}$Mo$_6$O$_{17}$[52], LaO$_{0.5}$F$_{0.5}$BiS$_2$[15], La$_2$IRu$_2$[53], and K$_2$Mo$_3$As$_3$[54], is also ruled out. Because both the crystal structure and the Fermi surfaces (see Figure S5) are clearly three-dimensional.

The reduction of $g$-factor can cause a violation of the Pauli limit. To elucidate this, a fit to the normal state magnetization data is carried out based on $\chi(T) = \chi_0 + C/(T - \theta)$, giving a temperature independent term of $\chi_0 = 1.73 \times 10^{-3}$ emu mol$^{-1}$. After subtracting the core diamagnetic contributions from Ti$^{4+}$, Ir$^{4+}$, and O$^{2-}$, we obtain an estimation of the Pauli paramagnetic susceptibility $\chi_P \sim 1.84 \times 10^{-3}$ emu mol$^{-1}$. Therefore, the Wilson ratio $R_W = \pi^2 k_B^2 \chi_P/(3\gamma\mu_B^2) \sim 3.9$. Such a large $R_W$ indicates the existence of strong electron correlation and/or a low-lying magnetic order[55, 56]. The strong electron correlation is further evidenced by the Kadowaki-Woods ratio $A/\gamma^2 = 9.0 \times 10^{-6}$ $\mu\Omega$ cm/(mJ mol$^{-1}$ K$^{-1}$)$^2$, which is comparable with those in heavy fermion systems (for instance, $1.0 \times 10^{-5}$ $\mu\Omega$ cm/(mJ mol$^{-1}$ K$^{-1}$)$^2$ for UPt$_3$[5] and CeCu$_2$Si$_2$[57]). It is also probably the cause for the enhancement of $\gamma$ compared with the theoretical value.

Finally, since the Pauli limit is based on the BCS weak coupling gap ratio of 1.76$k_B T_c$, a strong coupling can also enhance it. In our case, $\Delta_1/k_B T_c = 3.10$ is indeed beyond the BCS weak coupling limit. However, superconductors with similar coupling strength, such as the A15 compounds[41, 42] or SrPt$_3$P[43] do not necessarily host a large $H_{c2}$. We argue that the combination of strong-coupled superconductivity, substantial spin-orbit scattering, and the existence of strong electron correlation altogether result in the violation of Pauli limit.

To summarize, we report a detailed investigation of superconductivity in Ti$_4$Ir$_2$O. Being a strong-coupled superconductor with $T_c$ = 5.1 ∼ 5.7 K, Ti$_4$Ir$_2$O exhibits many intriguing properties. It has a large $\mu_0 H_{c2}(0)$ of 16.45 T, which is beyond the Pauli paramagnetic limit. A multigap superconductivity is evidenced by the temperature dependence of $H_{c2}$, the heat capacity measurements, and the DFT calculations. Moreover, a strong electron correlation in Ti$_4$Ir$_2$O, which is comparable with typical heavy fermion superconductors, is revealed by the large values of the Wilson ratio and the Kadowaki-Woods ratio, as well as the relatively large Sommerfeld coefficient ($\gamma$). These results make Ti$_4$Ir$_2$O a rare example to study the interplay between strong electron correlation and multigap superconductivity. We notice that these results also strongly suggest unconventional superconductivity. Questions like whether there are spin fluctuations, or a direct observation of the multigap feature through spectroscopy methods are still open, calling for further studies. Given the fact that there are numerous $\eta$-carbide type compounds, most of which are not systematically investigated for their transport properties, detailed studies in this structural family may flourish in the near future.

## 5. Conflict of interest

The authors declare that they have no conflict of interest.

## 6. Acknowledgments

The authors are grateful for the financial supports from the National Key Research and Development of China (Grant No. 2018YFA0704200, 2017YFA0302904, and 2018 YFA0305602), and the National Natural Science Foundation of China (No. 12074414, 12074002, and 11774402).

## CRediT authorship contribution statement

**Bin-Bin Ruan:** Conceptualization, Methodology, Formal analysis, Writing - Original draft preparation. **Meng-Hu Zhou:** Investigation, Writing - Review & Editing. **Qing-Song Yang:** Investigation. **Ya-Dong Gu:** Investigation. **Ming-Wei Ma:** Investigation. **Gen-Fu Chen:** Resources, Supervision, Writing - Review & Editing. **Zhi-An Ren:** Resources, Supervision, Writing - Review & Editing.

*Note*: During the preparation of the manuscript, we noticed the recent report (pulished in November, 2021) of superconductivity in Ti$_4$Ir$_2$O by Ma *et al.*[58]. The superconducting parameters determined there agreed fairly well with ours. However, the existence of strong correlation, as well as the multigap superconductivity, was overlooked in that study.

## References

[1] Gurevich, A.. To use or not to use cool superconductors? Nat Mater 2011;10(4):255–259. doi:10.1038/nmat2991.

[2] Hahn, S., Kim, K., Kim, K., Hu, X., Painter, T., Dixon, I., et al. 45.5-tesla direct-current magnetic field generated with a high-temperature superconducting magnet. Nature 2019;570(7762):496–499. doi:10.1038/s41586-019-1293-1.

[3] Stewart, G.. Superconductivity in the A15 structure. Physica C 2015;514:28–35. doi:10.1016/j.physc.2015.02.013; superconducting Materials: Conventional, Unconventional and Undetermined.

[4] Poole, C.K., Farach, H.A., Creswick, R.J.. Handbook of superconductivity. Elsevier; 1999.

[5] Stewart, G., Fisk, Z., Willis, J., Smith, J.. Possibility of Coexistence of Bulk superconductivity and Spin Fluctuations in UPt$_3$. Phys Rev Lett 1984;52(8):679–682. doi:10.1103/PhysRevLett.52.679.

[6] Hunte, F., Jaroszynski, J., Gurevich, A., Larbalestier, D.C., Jin, R., Sefat, A.S., et al. Two-band superconductivity in LaFeAsO$_{0.89}$F$_{0.11}$ at very high magnetic fields. Nature 2008;453(7197):903–905. doi:10.1038/nature07058.

[7] Bauer, E., Hilscher, G., Michor, H., Paul, C., Scheidt, E., Gribanov, A., et al. Heavy fermion superconductivity and magnetic order in noncentrosymmetric CePt3Si. Phys Rev Lett 2004;92(2). doi:10.1103/PhysRevLett.92.027003.

[8] Gupta, R., Ying, T., Qi, Y., Hosono, H., Khasanov, R.. Gap symmetry of the noncentrosymmetric superconductor W$_3$Al$_2$C. Phys Rev B 2021;103(17):174511. doi:10.1103/PhysRevB.103.174511.

[9] Bao, J.K., Liu, J.Y., Ma, C.W., Meng, Z.H., Tang, Z.T., Sun, Y.L., et al. Superconductivity in quasi-one-dimensional K$_2$Cr$_3$As$_3$ with significant electron correlations. Phys Rev X 2015;5(1):011013. doi:10.1103/PhysRevX.5.011013.

[10] Zhi, H., Imai, T., Ning, F., Bao, J.K., Cao, G.H.. NMR investigation of the quasi-one-dimensional superconductor K$_2$Cr$_3$As$_3$. Phys Rev Lett 2015;114(14):147004. doi:10.1103/PhysRevLett.114.147004.

[11] Tang, Z.T., Bao, J.K., Liu, Y., Sun, Y.L., Ablimit, A., Zhai, H.F., et al. Unconventional superconductivity in quasi-one-dimensional Rb$_2$Cr$_3$As$_3$. Phys Rev B 2015;91(2):020506. doi:10.1103/PhysRevB.91.020506.

[12] Cao, Y.C., Park, J.M., Watanabe, K., Taniguchi, T., Jarillo-Herrero, P.. Pauli-limit violation and re-entrant superconductivity in moiré graphene. Nature 2021;595(7868):526+. doi:10.1038/s41586-021-03685-y.

[13] Lu, Y., Takayama, T., Bangura, A.F., Katsura, Y., Hashizume, D., Takagi, H.. Superconductivity at 6 k and the violation of pauli limit in Ta$_2$Pd$_x$S$_5$. J Phys Soc Jpn 2014;83(2):023702. doi:10.7566/JPSJ.83.023702.

[14] Okuda, K., Kitagawa, M., Sakakibara, T., Date, M.. Upper critical field measurements up to 600 kG in PbMo$_6$S$_8$. J Phys Soc Jpn 1980;48(6):2157–2158. doi:10.1143/JPSJ.48.2157.

[15] Chan, Y.C., Yip, K.Y., Cheung, Y.W., Chan, Y.T., Niu, Q., Kajitani, J., et al. Anisotropic two-gap superconductivity and the absence of a Pauli paramagnetic limit in single-crystalline LaO0.5F0.5BiS2. Phys Rev B 2018;97(10). doi:10.1103/PhysRevB.97.104509.

[16] Taylor, A., Sachs, K.. A new complex of η-carbide. Nature 1952;169(4297):411. doi:10.1038/169411a0.

[17] Kuo, K.. The Formation of η-carbides. Acta Metall 1953;1(3):301–304. doi:10.1016/0001-6160(53)90103-5.

[18] Newsam, J., Jacobson, A., McCandlish, L., Polizzotti, R.. The structures of the η-carbides Ni$_6$Mo$_6$C, Co$_6$Mo$_6$C, and Co$_6$Mo$_6$C$_2$. J Solid State Chem 1988;75(2). doi:10.1016/0022-4596(88)90170-3.

[19] Jeitschko, W., Holleck, H., Nowotny, H., Benesovsky, F.. Phasen mit aufgefülltem Ti$_2$Ni-typ. Monatsh Chem 1964;95(3):1004–1006. doi:10.1007/BF00908814.

[20] Dubrovinskaia, N., Dubrovinsky, L., Saxena, S., Selleby, M., Sundman, B.. Thermal expansion and compressibility of Co$_6$W$_6$C. J Alloys Compd 1999;285(1-2):242–245. doi:10.1016/S0925-8388(98)00932-3.

[21] Zavalii, I., Verbovytskyi, Y., Berezovets, V., Shtender, V., Pecharsky, V., Lyutyi, P.. Synthesis, structure, and hydrogen-sorption properties of (Ti,Zr)$_4$Ni$_2$n$_x$ subnitrides. Mater Sci 2017;53(3):306–315. doi:10.1007/s11003-017-0076-9.

[22] Nagai, M., Zahidul, A.M., Matsuda, K.. Nano-structured nickel–molybdenum carbide catalyst for low-temperature water-gas shift reaction. Appl Catal, A 2006;313(2):137–145. doi:10.1016/j.apcata.2006.07.006.

[23] Ku, H., Johnston, D.. New superconducting ternary transition metal compounds with the e93-type structure. Chin J Phys 1984;22(1):59–64.

[24] Ma, K., Lago, J., von Rohr, F.O.. Superconductivity in the $\eta$-carbide-type oxides Zr$_4$Rh$_2$O$_x$. J Alloys Compd 2019;796:287–292. doi:10.1016/j.jallcom.2019.04.318.

[25] Ma, K., Gornicka, K., Lefèvre, R., Yang, Y., Rønnow, H.M., Jeschke, H.O., et al. Superconductivity with high upper critical field in the cubic centrosymmetric $\eta$-carbide nb4rh2c1-$\delta$. ACS Materials Au 2021;doi:10.1021/acsmaterialsau.1c00011.

[26] Matthias, B., Geballe, T., Compton, V.. Superconductivity. Rev Mod Phys 1963;35(1):1–&. doi:10.1103/RevModPhys.35.1.

[27] Zegler, S.. Superconductivity in zirconium-rhodium alloys. J Phys Chem Solids 1969;26(8):1347–&. doi:10.1016/0022-3697(65)90117-4.

[28] Toby, B.. Expgui, a graphical user interface for gsas. J Appl Crystallogr 2001;34(2):210–213. doi:10.1107/S0021889801002242.

[29] Ruan, B.B., Yang, Q.S., Zhou, M.H., Chen, G.F., Ren, Z.A.. Superconductivity in a new T$_2$-phase Mo$_5$GeB$_2$. J Alloys Compd 2021;868:159230. doi:10.1016/j.jallcom.2021.159230.

[30] Prozorov, R., Kogan, V.G.. Effective demagnetizing factors of diamagnetic samples of various shapes. Phys Rev Appl 2018;10(1):014030. doi:10.1103/PhysRevApplied.10.014030.

[31] Giannozzi, P., Baroni, S., Bonini, N., Calandra, M., Car, R., Cavazzoni, C., et al. QUANTUM ESPRESSO: a modular and open-source software project for quantum simulations of materials. J Phys: Condens Matter 2009;21(39). doi:10.1088/0953-8984/21/39/395502.

[32] Giannozzi, P., Andreussi, O., Brumme, T., Bunau, O., Nardelli, M.B., Calandra, M., et al. Advanced capabilities for materials modelling with QUANTUM ESPRESSO. J Phys: Condens Matter 2017;29(46):465901. URL: https://doi.org/10.1088/1361-648x/aa8f79. doi:10.1088/1361-648x/aa8f79.

[33] Giannozzi, P., Baseggio, O., Bonfa, P., Brunato, D., Car, R., Carnimeo, I., et al. QUANTUM ESPRESSO toward the exascale. J Chem Phys 2020;152(15). doi:10.1063/5.0005082.

[34] Perdew, J.P., Ruzsinszky, A., Csonka, G.I., Vydrov, O.A., Scuseria, G.E., Constantin, L.A., et al. Restoring the density-gradient expansion for exchange in solids and surfaces. Phys Rev Lett 2008;100(13). doi:10.1103/PhysRevLett.100.136406.

[35] Dal Corso, A.. Pseudopotentials periodic table: From H to Pu. Comput Mater Sci 2014;95:337–350. doi:10.1016/j.commatsci.2014.07.043.

[36] Otero-de-la Roza, A., Johnson, E.R., Luana, V.. CRITIC2: A program for real-space analysis of quantum chemical interactions in solids. Comput Phys Commun 2014;185(3):1007–1018. doi:10.1016/j.cpc.2013.10.026.

[37] Fuhr, J.D., Roura-Bas, P., Aligia, A.A.. Maximally localized Wannier functions for describing a topological phase transition in stanene. Phys Rev B 2021;103(3). doi:10.1103/PhysRevB.103.035126.

[38] Pizzi, G., Vitale, V., Arita, R., Bluegel, S., Freimuth, F., Geranton, G., et al. Wannier90 as a community code: new features and applications. J Phys: Condens Matter 2020;32(16). doi:10.1088/1361-648X/ab51ff.

[39] Okamoto, H., Massalski, T., et al. Binary alloy phase diagrams. Materials Park, OH, USA: ASM International; 1990.

[40] Xu, C., Wang, H., Tian, H., Shi, Y., Li, Z.A., Xiao, R., et al. Superconductivity in an intermetallic oxide Hf$_3$Pt$_4$Ge$_2$O. Chin Phys B 2021;doi:10.1088/1674-1056/abfb53.

[41] Carbotte, J.. Properties of Boson-Exchange Superconductors. Rev Mod Phys 1990;62(4):1027–1157. doi:10.1103/RevModPhys.62.1027.

[42] Fisk, Z., Webb, G.. Saturation of the high-temperature normal-state electrical resistivity of superconductors. Phys Rev Lett 1976;36(18):1084. doi:10.1103/PhysRevLett.36.1084.

[43] Takayama, T., Kuwano, K., Hirai, D., Katsura, Y., Yamamoto, A., Takagi, H.. Strong Coupling Superconductivity at 8.4 K in an Antiperovskite Compound SrPt3P. Phys Rev Lett 2012;108(23). doi:10.1103/PhysRevLett.108.237001.

[44] Mott, N.F., Jones, H., Jones, H., Jones, H.. The theory of the properties of metals and alloys. Courier Dover Publications; 1958.

[45] Mott, N.F.. Electrons in transition metals. Adv Phys 1964;13(51):325–422. doi:10.1080/00018736400101041.

[46] Werthamer, N., Helfand, E., PC, H.. Temperature and purity dependence of the superconducting critical field, Hc2. II. Phys Rev 1966;147(1):288. doi:10.1103/PhysRev.147.295.

[47] Bud'Ko, S., Petrovic, C., Lapertot, G., Cunningham, C., Canfield, P., Jung, M., et al. Magnetoresistivity and H$_{c2}$(T) in MgB$_2$. Phys Rev B 2001;63(22):220503. doi:10.1103/PhysRevB.63.220503.

[48] Suderow, H., Tissen, V., Brison, J., Martínez, J., Vieira, S.. Pressure induced effects on the fermi surface of superconducting 2H-NbSe$_2$. Phys rev lett 2005;95(11):117006. doi:{10.1103/PhysRevLett.95.117006}.

[49] De Faria, L.R., Ferreira, P.P., Correa, L.E., Eleno, L.T., Torikachvili, M.S., Machado, A.J.. Possible multiband superconductivity in the quaternary carbide YRe$_2$SiC. Supercond Sci Technol 2021;34(6):065010. doi:10.1088/1361-6668/abf7cf.

[50] Gurevich, A.. Enhancement of the upper critical field by nonmagnetic impurities in dirty two-gap superconductors. Phys Rev B 2003;67(18). doi:10.1103/PhysRevB.67.184515.

[51] McMillan, W.L.. Transition temperature of strong-coupled superconductors. Phys Rev 1968;167:331. doi:10.1103/PhysRev.167.331.

[52] Mercure, J.F., Bangura, A.F., Xu, X., Wakeham, N., Carrington, A., Walmsley, P., et al. Upper Critical Magnetic Field far above the Paramagnetic Pair-Breaking Limit of Superconducting One-Dimensional Li$_{0.9}$Mo$_6$O$_{17}$ Single Crystals. Phys Rev Lett 2012;108(18). doi:10.1103/PhysRevLett.108.187003.

[53] Ishikawa, H., Wedig, U., Nuss, J., Kremer, R.K., Dinnebier, R., Blankenhorn, M., et al. Superconductivity at 4.8 K and Violation of Pauli Limit in La2IRu2 Comprising Ru Honeycomb Layer. Inorg Chem 2019;58(19):12888–12894. doi:10.1021/acs.inorgchem.9b01825.

[54] Mu, Q.G., Ruan, B.B., Zhao, K., Pan, B.J., Liu, T., Shan, L., et al. Superconductivity at 10.4 k in a novel quasi-one-dimensional ternary molybdenum pnictide K$_2$Mo$_3$As$_3$. Sci Bull 2018;63(15):952–956. doi:10.1016/j.scib.2018.06.011.

[55] Anand, V.K., Adroja, D.T., Hillier, A.D.. Ferromagnetic cluster spin-glass behavior in PrRhSn3. Phys Rev B 2012;85(1). doi:10.1103/PhysRevB.85.014418.

[56] Tong, P., Sun, Y.P., Zhu, X.B., Song, W.H.. Strong electron-electron correlation in the antiperovskite compound GaCNi3. Phys Rev B 2006;73(24). doi:10.1103/PhysRevB.73.245106.

[57] Steglich, F., Aarts, J., Bredl, C., Lieke, W., Meschede, D., Franz, W., et al. Superconductivity In The Presence Of Strong Pauli Paramagnetism - CeCu$_2$Si$_2$. Phys Rev Lett 1979;43(25):1892–1896. doi:10.1103/PhysRevLett.43.1892.

[58] Ma, K., Lefèvre, R., Gornicka, K., Jeschke, H.O., Zhang, X., Guguchia, Z., et al. Group-9 Transition-Metal Suboxides Adopting the Filled-Ti$_2$Ni Structure: A Class of Superconductors Exhibiting Exceptionally High Upper Critical Fields. Chem Mater 2021;doi:10.1021/acs.chemmater.1c02683.

# Superconductivity with a Violation of Pauli Limit and Evidences for Multigap in $\eta$-Carbide type Ti$_4$Ir$_2$O

Bin-Bin Ruan [a, b, *], Meng-Hu Zhou [a, b], Qing-Song Yang [b, c], Ya-Dong Gu [b, c], Ming-Wei Ma [b, c], Gen-Fu Chen [b, c], Zhi-An Ren [b, c, #]

[a] Songshan Lake Materials Laboratory, Dongguan, Guangdong 523808, China
[b] Institute of Physics and Beijing National Laboratory for Condensed Matter Physics, Chinese Academy of Sciences, Beijing 100190, China
[c] School of Physical Sciences, University of Chinese Academy of Sciences, Beijing 100049, China

[*] Corresponding author. E-mail: bbruan@mail.ustc.edu.cn (B-B. Ruan).
[#] Corresponding author. E-mail: renzhian@iphy.ac.cn (Z-A. Ren).

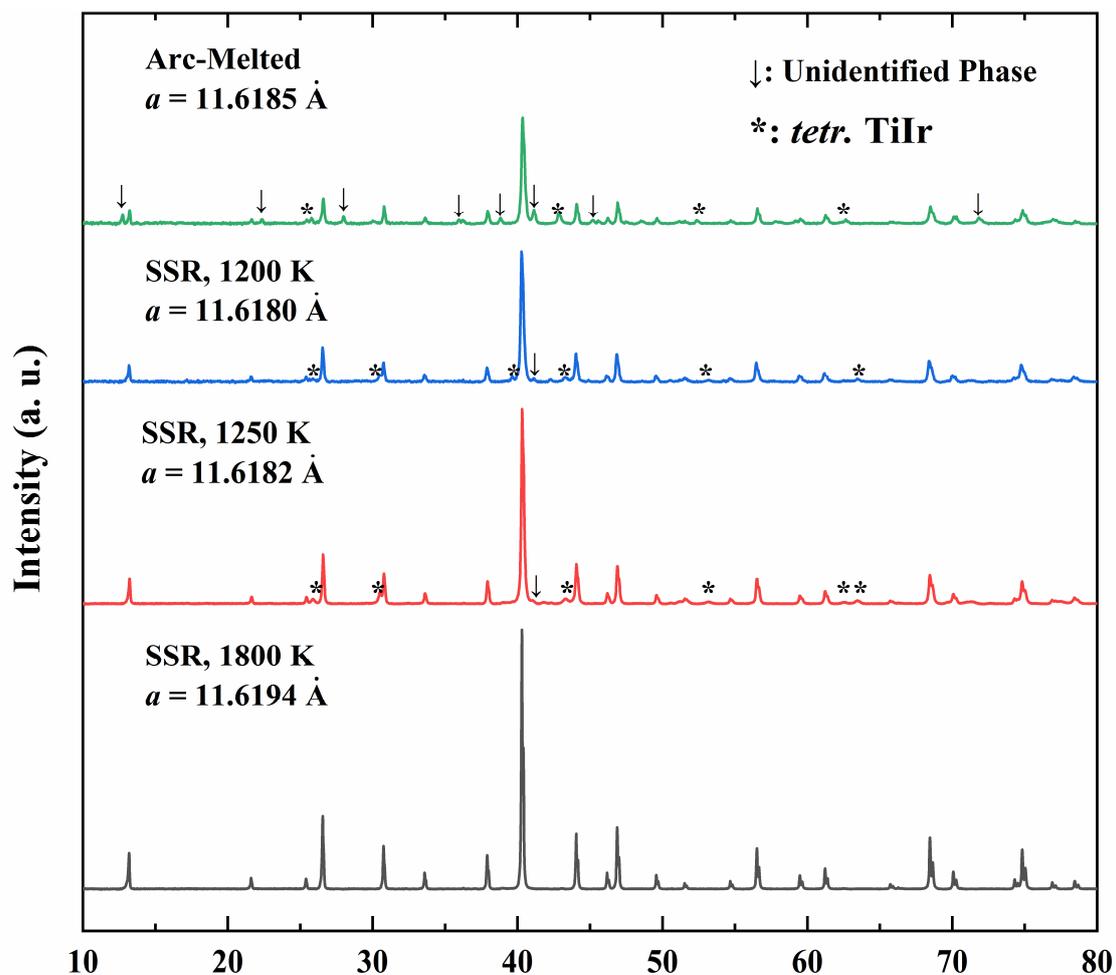

**Figure S1**: Representative XRD patterns of $Ti_4Ir_2O$ samples prepared with different methods. From bottom to top: SSR (Solid state reaction) at 1800 K, SSR at 1250 K, SSR at 1200 K, and arc-melted samples. SSR at 1800 K produced single-phase sample, while the other three methods produced multi-phase samples. The lattice parameter *a* is slightly different too. *Tetr.* TiIr stands for tetragonal TiIr (space group $P4/mmm$). "↓" indicates an unidentified phase.

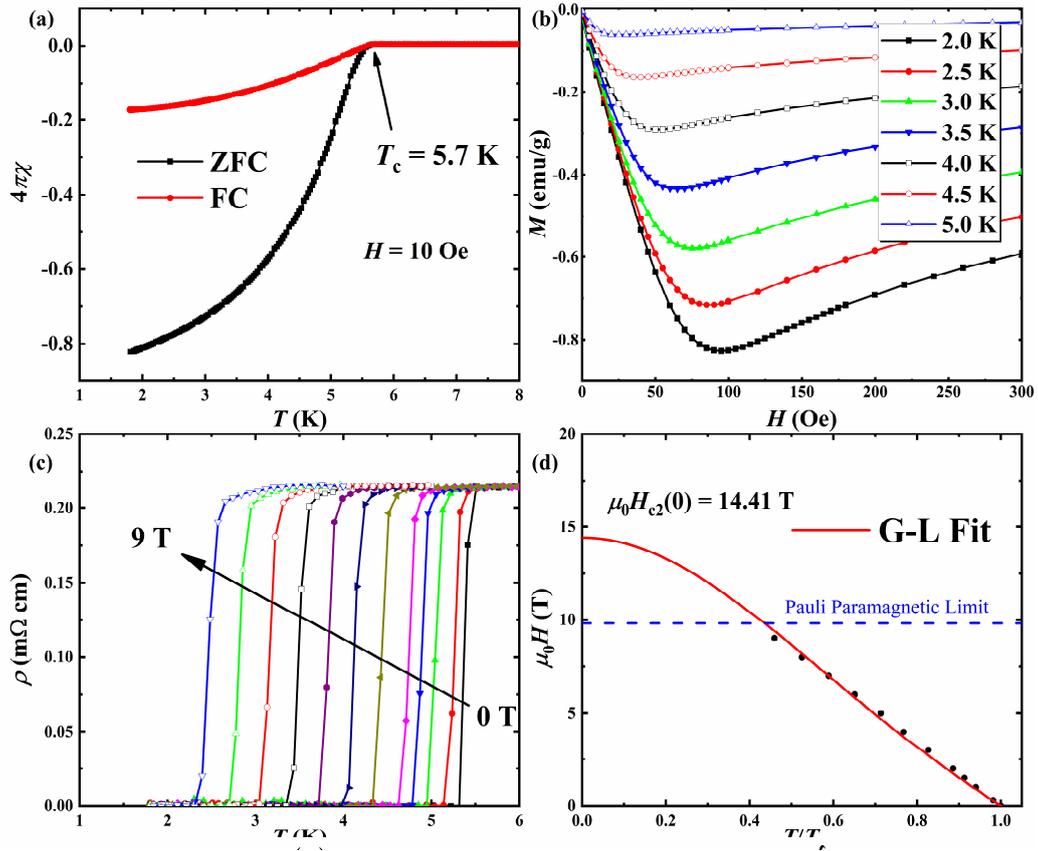

**Figure S2**: Electrical and magnetic properties of $Ti_4Ir_2O$ prepared by SSR at 1200 K. Notice that although $T_c$ of this sample is somewhat higher than the one shown in the main text, the transition width is larger, and the upper critical field is lower (but still above the Pauli limit), indicating a less good sample quality. (a) DC magnetic susceptibility under 10 Oe from 1.8 K to 8 K. (b) Isothermal magnetization curves at different temperatures. (c) Superconducting transition under different magnetic fields. (d) Upper critical fields at different temperatures, and its fit by G–L relation.

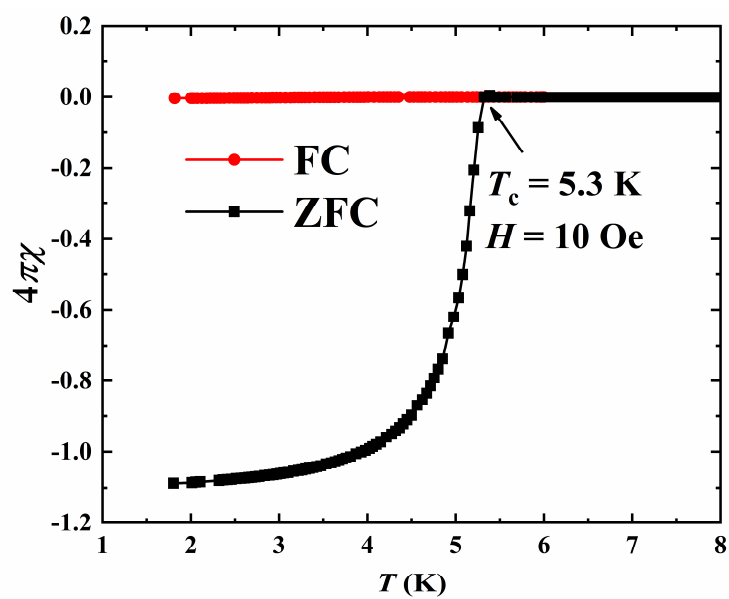

**Figure S3**: DC magnetic susceptibility of $Ti_4Ir_2O$ prepared by arc-melting method.

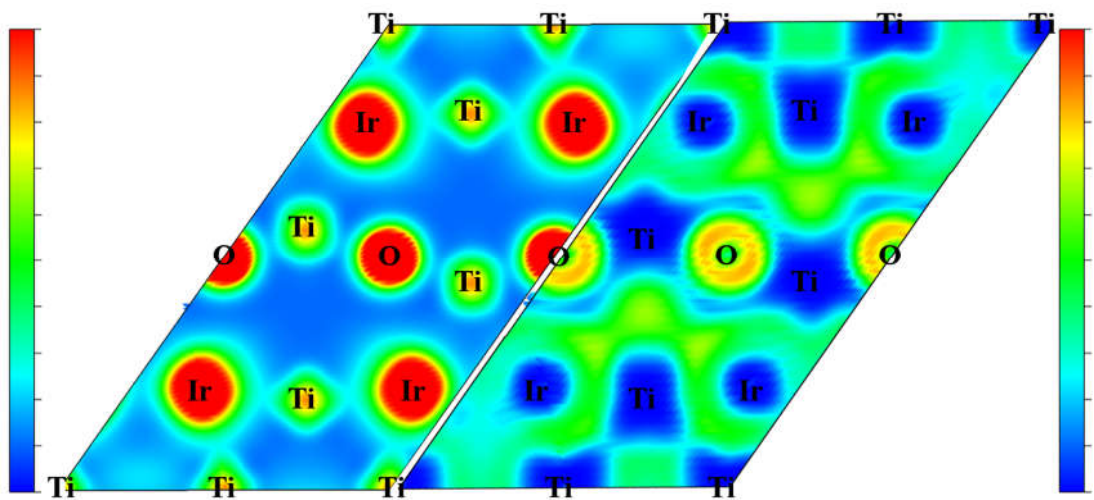

**Figure S4**: Charge density (left) and electron localization function (ELF) (right) of $Ti_4Ir_2O$ calculated with SOC. The charge density is bound between 0 and 0.2 $e/a_0^3$ ($a_0$ is the Bohr radius), while ELF is bound between 0 and 1.

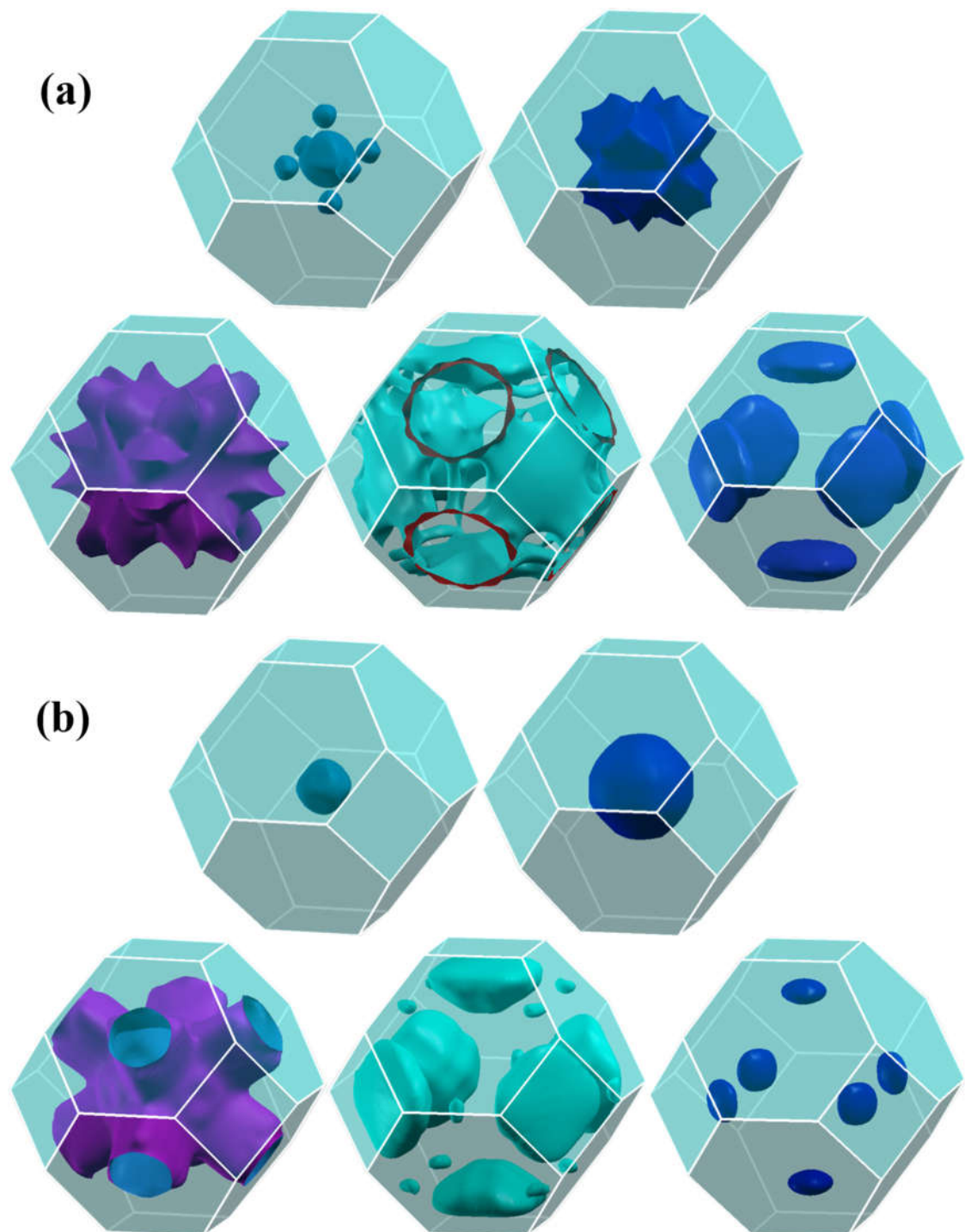

**Figure S5**: Fermi surfaces of $Ti_4Ir_2O$ (a) without SOC and (b) with SOC.